\documentclass[twocolumn,superscriptaddress,amssymb,amsmath,nobibnotes,aps,prd]{revtex4}

\usepackage{color}
\usepackage{graphicx}
\usepackage{comment}

\newcommand{\ee}{{\mathrm e}}
\newcommand{\ii}{{\mathrm i}}

\newcommand{\zz}{{\mathbb Z}}
\newcommand{\Tr}{\operatorname{Tr}}

\newcommand{\ndif}[1][]{{\mathrm{d}^{#1} \!}}
\newcommand{\dif}[2][]{{\ndif[#1] #2 \,}}

\begin{document}

\title{The entanglement and relative entropy of a chiral fermion on the torus}

\author{Pascal Fries}
\email{pascal.fries@physik.uni-wuerzburg.de}
\affiliation{Fakult\"at f\"ur Physik und Astronomie, Julius-Maximilians Universit\"at
  W\"urzburg, Am Hubland, 97074 W\"urzburg, Germany}
\author{Ignacio A. Reyes}
\email{ignacio.reyes@aei.mpg.de}
\affiliation{Max-Planck-Institut f\"ur Gravitationsphysik, Am M\"uhlenberg 1, 14476
  Potsdam, Germany}

\date{\today}

\begin{abstract}
  We derive the entanglement entropy of chiral fermions on the circle at arbitrary
  temperature. The spin-sector contribution depends only on the total length of the
  entangling region, regardless of the configuration of the intervals. Thus three-partite
  information provides a global indicator for the spin boundary conditions. Together with
  the modular Hamiltonian, our results provide a systematic way of obtaining relative
  entropy on the torus.
\end{abstract}

\maketitle

\section{Introduction}
Entanglement is perhaps the most characteristic feature of quantum theory. Although there
exists many measures for it, the most important one is the \emph{entanglement
  entropy}. Given a global state defined via a density matrix $\rho$ and a subsystem $A$
of it, the reduced density matrix on $A$ is defined as $\rho_A:=\text{tr}_B(\rho)$, where
$B$ is the complement of $A$. Then, the entanglement entropy between $A$ and $B$ is simply
the von Neumann entropy of the subsystem,
\begin{equation}
  \label{SA}
  S(A) =-\Tr_A \big[\rho_A\log \rho_A \big].
\end{equation}

In practice however, computing the logarithm in~\eqref{SA} is generically very
challenging. Therefore, it is customary in QFT to resort to the \emph{replica trick}. This
consists of first computing a closely related quantity - the R\'enyi entropy for integer
index $n$:
\begin{equation}
  \label{Sn}
  S_n(A)=\frac{1}{1-n}\log \Tr \rho_A^n.
\end{equation}

This requires to compute the partition function of the theory on the $n-$fold cover of the
manifold of interest, where the $n$ copies are glued cyclically along the entangling
region $A$. In order for this strategy to work, it is crucial that we are able to find the
analytic continuation of the R\'enyi entropy~\eqref{Sn} away from integer $n$ and take the
limit $n\rightarrow 1$. To the best of our knowledge such analytic continuation for the
entanglement entropy of free fermions on the torus (finite temperature on the circle) is
not known for arbitrary torus modulus.

In this paper, we derive the entanglement entropy of the chiral fermion on the torus via a
different route - by computing the entropy directly from the resolvent. This avoids the
need of finding the analytic continuation of the R\'enyi entropies. For the free fermion
on the circle at finite temperature, the R\'enyi entropies for integer $n$ where first
computed in~\cite{Azeyanagi:2007bj} for a single interval, and in~\cite{Herzog:2013py} for
multiple intervals. The analytic continuation to the entanglement entropy however was
obtained for the high and low temperature expansions only.

The resolvent for the chiral fermion on the torus was recently obtained
in~\cite{Fries:2019ozf} and also in~\cite{Blanco:2019xwi}. As direct applications of the
resolvent and modular Hamiltonian, we study entanglement entropy, mutual information,
three-partite information and relative entropy. Interestingly, while three-partite
information vanishes for fermions in the plane~\cite{Casini:2008wt}, it is generally
non-zero on the torus. Instead, we find that it is a global indicator of the boundary
conditions, in the sense that its value does not depend on the precise configuration of
the intervals involved.

The paper is organised as follows. In section~\ref{Resolvent} we review the resolvent
method, the main tool that we will apply to later computations. In section~\ref{EE} we go
straight to the calculation of the entanglement entropy, and discuss the mutual
information. In section~\ref{Relative} we consider relative entropy and illustrate some
properties taking a small interval expansion. We summarise and discuss our results in
section~\ref{Discussion}.

\section{Review of the resolvent}
\label{Resolvent}
Let us quickly summarise the definition of the \emph{resolvent} and the mains steps
involved in its derivation for the fermions on the torus~\cite{Fries:2019ozf}. A global
state $\rho$ for the system on a generic Riemann surface determines the equal-time
correlation function $G(x,y):=\langle \psi(x)\psi^\dag(y)\rangle$. The \emph{resolvent} of
the correlator is defined via:
\begin{equation}
  \label{Res}
  R(\xi):= [G+\xi-1/2]^{-1}
\end{equation}
where we always leave the space-time dependence of $R$ implicit. In~\eqref{Res}, the
inverse must be understood in the sense of distributions, i.e. as integrated agains
regular test functions. Eq.~\eqref{Res} thus involves solving an integral equation for
$R$.

One of the central results of~\cite{Fries:2019ozf} is the resolvent of a single chiral
fermion on the torus for an arbitrary number $p$ of disjoint intervals
$A=\cup_{j=1}^p (a_j,b_j)$, with both $\nu=2$ (R-NS) and $\nu=3$ (NS-NS) boundary
conditions. We shall not focus on its derivation, but rather quickly review its structure
and explain its properties. Explicitly it is given by
\begin{equation}
  \label{R}
  R(\xi)=\frac{\delta(x-y)}{\xi-1/2} - \frac{F(x,y;\xi)}{(\xi-1/2)^2}
\end{equation}
where $x,y\in A$ are any two points on the entangling region, and
\begin{equation}
  \label{F}
  F(x,y;\xi)=\frac{\xi-1/2}{\xi+1/2} G_\nu(x-y|\tau,Lh) \bigg[\frac{\Omega(x|\tau)}{\Omega(y|\tau)}\bigg]^{\ii h}
\end{equation}
where each term is defined as
\begin{align}
  h
  &= \frac{1}{2\pi}\log \frac{\xi+1/2}{\xi-1/2} \\
  G_\nu(z|\tau,\mu)
  &= \frac{\eta^3(\tau)}{\ii\vartheta_1(z|\tau)} \frac{\vartheta_\nu(z-\ii \mu|\tau)}{\vartheta_\nu(-\ii\mu|\tau)}, \label{Gzmu} \\
  \Omega(x|\tau)
  &= -\prod_{j=1}^p \frac{\vartheta_1(x-a_j|\tau)}{\vartheta_1(x-b_j|\tau)}. \label{Omega}
\end{align}

Here, $L=\sum_j |b_j-a_j|$ is the total length of all intervals and our conventions for
the Dedekind eta and the Jacobi elliptic functions are described in the Appendix. Note the
appearance of $G_\nu (z|\tau,Lh)$, which is the propagator on the torus with a chemical
potential $\mu = Lh$ as we will show in Sec.~\ref{ChemicalPotential}. As usual, we fix
by convenience the periods of the torus to be $1,\tau$ -- physics on the torus depends
only on $\tau$ and we can always recover the result for a different spatial circle by
rescaling. We shall restrict to purely imaginary modulus $\tau=\ii\beta$ where $\beta$ is
the inverse temperature, keeping in mind that the general case can be recovered by
analytic continuation.

The knowledge of the resolvent was further used to derive the modular Hamiltonian for an
arbitrary number of intervals on the torus. The modular Hamiltonian $K$ of a density
matrix is defined via
\[
  \rho = \frac{\ee^{-K}}{\Tr \ee^{-K}}
\]
and has found numerous applications in many body quantum
systems~\cite{Chandran:2011sdf,Vanderstraeten:2017unh,Zhu:2018wsx}, quantum
information~\cite{Li:2008kjh,Cirac:2011iuz,Dalmonte:2017bzm}, quantum field
theory~\cite{Casini:2009vk,Casini:2008cr,Cardy:2016fqc,Faulkner:2016mzt,Klich:2017qmt,Arias:2018tmw},
modular theory~\cite{Bisognano:1976za,Lashkari:2015dia} and the AdS/CFT
correspondence~\cite{Casini:2011kv,Blanco:2013joa,Faulkner:2013ica,Jafferis:2014lza,Belin:2018juv,Chen:2018rgz}.

The modular Hamiltonian for the chiral fermion on the torus exhibits a surprisingly
interesting structure. Even for a single interval, the modular flow couples any given
point to an infinite but discrete set of other points. These accumulate near the
boundaries of the interval, where their contribution becomes increasingly damped or
``red-shifted'' as they approach the endpoints. In the limit of zero temperature, these
points `condense' regularly in the interval, giving rise to continuous
non-locality~\cite{Fries:2019ozf}.

\section{Entanglement entropy}
\label{EE}
We now determine the entanglement entropy on the torus of an arbitrary set of disjoint
intervals. As mentioned above, once the resolvent is obtained, the entanglement entropy
follows by a trace formula~\cite{Casini:2009vk}:
\begin{align*}
S=-\Tr \int_{1/2}^\infty \!\!\dif\xi \left[(\xi-1/2) [R(\xi)-R(-\xi)] - \frac{2\xi}{\xi+1/2} \right] 
\end{align*}

Replacing the resolvent~\eqref{R} one finds
\begin{align}
  S
  &= \int_{A} \dif x \lim_{y\to x} \int_{1/2}^\infty \dif\xi \nonumber\\
  &\hspace{1cm}\times (\xi-1/2) \left[\frac{F(\xi)}{(\xi-1/2)^2} - \frac{F(-\xi)}{(\xi+1/2)^2} \right]. \label{Slim}
\end{align}

Next, we substitute $F$ from~\eqref{F}, and the integrand becomes
\[
  \frac 1{\xi+1/2} \lim_{y\to x} \bigg[ G_\nu(x-y|\tau,Lh)
  \bigg[\frac{\Omega(x|\tau)}{\Omega(y|\tau)} \bigg]^{\ii h}\!\!\!\! - (h\to -h) \bigg].
\]

In the limit $y\to x$, $G(x-y|\tau,Lh)$ diverges like the UV propagator $1/2\pi\ii(x-y)$,
but this leading divergence cancels in the above difference, leaving a well defined
expression. To extract the finite contribution, we use the Laurent series~\eqref{Laurent3}
and~\eqref{Laurent2} provided in the Appendix. Then, the integrand in~\eqref{Slim} is
\begin{align}
  \frac 1{\xi+1/2} \bigg[\frac{h}{\pi} \partial_x \log \Omega(x|\tau) + 2\Sigma_\nu (\Lambda|\tau) \bigg] \,. \label{limit}
\end{align}

Here, the first term is position-dependent but identical for each spin sector, whereas the
second term is spin-dependent but spacially constant. It is given by a Laurent expansion
\begin{align}\label{}
  \Sigma_3(\Lambda|\tau)
  &= \sum_{\substack{k\geq 1\\k \text{ odd}}} \bigg[\frac{q^k}{\Lambda^{-1}+q^k}-\frac{q^k}{\Lambda+q^k}\bigg],\\
  \Sigma_2(\Lambda|\tau)
  &= \frac 12\frac{\Lambda-1}{\Lambda+1}+ \!\!\sum_{\substack{k\geq 2\\k \text{ even}}} \bigg[\frac{q^k}{\Lambda^{-1}+q^k}-\frac{q^k}{\Lambda+q^k}\bigg],\label{sigma2}
\end{align}
where we used the convenient variable
\[
  \Lambda=\ee^{2 \pi L h} = \bigg[\frac{\xi+1/2}{\xi-1/2}\bigg]^L
\]
and the nome $q=\ee^{\ii\pi\tau}$.

Thus, we learn from~\eqref{limit} that the entanglement entropy decomposes into a
spin-independent part $S^{(0)}$ and a spin-dependent one $S^{(\nu)}$
\[
  S = S^{(0)}+S^{(\nu)}
\]
In the next subsections we consider them separately.

\subsection{Spin-independent entropy}
We start by rederiving the known results from~\cite{Herzog:2013py,Azeyanagi:2007bj}. Let us
start by focusing on the first contribution in~\eqref{limit}, which does not depend on the
boundary conditions. Introducing it back into~\eqref{Slim}, the two integrals decouple as
\[
  S^{(0)} = \frac{1}{2\pi^2} \int_{1/2}^\infty \!\!\dif\xi
  \frac{\log \frac{\xi+1/2}{\xi-1/2}}{\xi+1/2} \int_A \!\dif x \partial_x \log \Omega(x|\tau).
\]

We see that only boundary terms from the spacial integral contribute, one per each
endpoint of the intervals. This gives the final result for the spin-independent entropy
\begin{align}
  S^{(0)}
  &=\frac{1}{6} \bigg[\sum_{i,j} \log \left| \vartheta_1(b_i-a_j|\tau) \right| \nonumber \\
  &\hspace{1.5cm}- \sum_{i<j} \log \left| \vartheta_1(b_i-b_j|\tau) \vartheta_1(a_i-a_j|\tau) \right| \nonumber \\
  &\hspace{3cm}- p \log |\vartheta_1(\epsilon|\tau)|\bigg], \label{S0CH}
\end{align}
which takes a form very reminiscent of the one described in~\cite{Casini:2009vk}, but with
respect to the Jacobi $\vartheta_1$ function. Note that we introduced a UV regulator
$\epsilon>0$, integrating only within $(a_i+\epsilon,b_i-\epsilon)$, to avoid the infinite
pileup of entanglement at the endpoints of each interval.

\subsection{Spin-dependent entropy}
In this section we derive our first new result: the contribution to entanglement entropy
that does depend on the choice of boundary conditions. As mentioned above, we restrict to
the sectors $\nu=2,3$.  Let us start with $\nu=3$ or (NS-NS) which has no zero
mode. Plugging the second term of~\eqref{limit} back into~\eqref{Slim}, we have
\[
  S^{(3)} = 2L \int_{1/2}^\infty  \frac{\dif \xi}{\xi+1/2} \Sigma_3(\Lambda|\tau)
\]

Notice that we performed the trivial integral over the entangling region right away to
yield a global prefactor of $L$, the total length of the regions. The remaining integral
is however much more challenging and finding a closed form goes beyond the scope of this
paper. In order to bring it into a more explicit form, we can change the variable
of integration to $\Lambda$, so that
\begin{align}
  S^{(3)}
  &= 2\sum_{\substack{k\geq 1\\ \text{ odd}}} \int_1^\infty \frac{d\Lambda }{\Lambda(\Lambda^{1/L}-1)} \nonumber \\
  &\hspace{2cm}\times \bigg[\frac{q^k}{\Lambda^{-1}+q^k} - \frac{q^k}{\Lambda+q^k}\bigg]. \label{S3}
\end{align}
Notice the integral is completely regular since $L<1$, as we fixed the length of the
spatial circle to unity. We have found no analytic expression for~\eqref{S3}, but it can
easily be dealt with numerically.

The case $\nu=2$ or (R-NS) is special, for it has a zero mode. This manifests in the
presence of the additional term in~\eqref{sigma2}, and gives an entropy of
\begin{align}
  S^{(2)}
  &=\int_1^\infty \frac{\dif\Lambda }{\Lambda(\Lambda^{1/L}-1)}\nonumber\\
  &\hspace{.5cm} \times\bigg[\frac{\Lambda-1}{\Lambda+1} +
    2\sum_{\substack{k\geq 2\\ \text{ even}}}\frac{q^k}{\Lambda^{-1}+q^k} - \frac{q^k}{\Lambda+q^k} \bigg],\label{S2}
\end{align}
which can again be computed numerically. 

The main result of this section is that the spin-dependent contribution to the entropy
depends only on the total length of the entangling region -- a feature also present
for the R\'enyi entropies~\cite{Herzog:2013py}.

\subsection{Mutual and three-partite information}
The mutual information between two disjoint intervals $A$ and $B$ is another important
information theory quantity. It it a measure of the correlation between two distributions,
here the two reduced density matrices $\rho_A$ and $\rho_B$, and is defined via
\[
  I(A,B) := S(A) + S(B) - S(A B)
\]

\begin{figure}[h]
\begin{center}
\includegraphics[width=7.9cm]{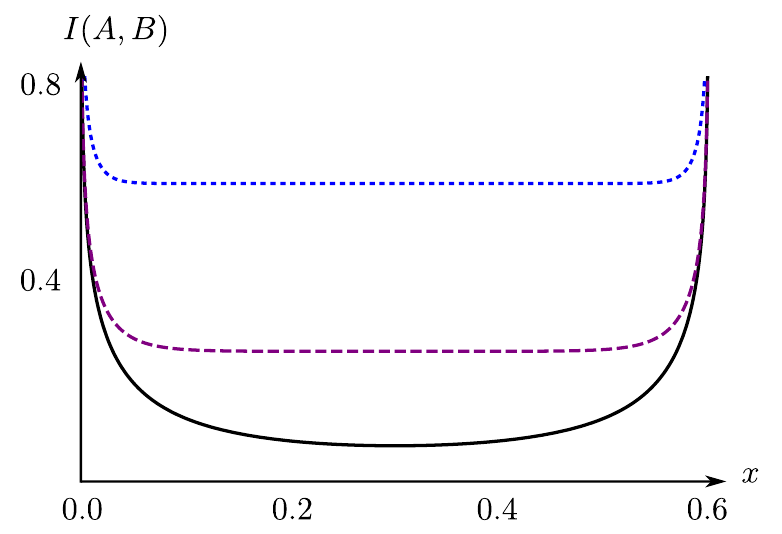} 
\end{center}
\caption{Spin-independent part of the mutual information between the intervals $A=(0,0.2)$
  and $B=(0.2+x,0.4+x)$ as we vary their separation $x$, for $\beta=3$ (solid black),
  $\beta=0.16$ (dashed purple), $\beta=0.07$ (dotted blue).}
\label{figmutual}
\end{figure}

From \eqref{S0CH}, the spin-independent part of the mutual information between two
non-intersecting intervals $A =(a_1,b_1)$ and $B =(a_2,b_2)$ is explicitly given by
\begin{equation}
  \label{IAB}
  I(A,B) = \frac{1}{6} \log \bigg|
  \frac{\vartheta_1(a_2-a_1|\tau)\vartheta_1(b_2-b_1|\tau)}
  {\vartheta_1(b_2-a_1|\tau) \vartheta_1(b_1-a_2|\tau)}\bigg|.
\end{equation}
In Fig.~\ref{figmutual} we plot this expression for different temperatures, as we
continuously vary the separation between the two intervals.

Remarkably, as pointed out in the previous section, the spin-dependent part to the entropy
depends on the entangling region only via its total length $L$. Therefore, its
contribution to the mutual information does not depend on the distance between the
intervals but only on their size
\[
  S^{(\nu)}(A) + S^{(\nu)}(B) -S^{(\nu)}(AB) = f(L_A,L_B).
\]

This is one the main results of this paper: mutual information is insensible to the
fermion boundary conditions up to a \emph{global} term. This was also shown
in~\cite{Herzog:2013py} for the R\'enyi mutual information $I_n(A,B)$ for
$n\geq 2\in \mathbb Z$.

This property, together with the spin-independent entropy~\eqref{S0CH} implies that on the
torus---in contrast to the plane~\cite{Casini:2008wt}---the three-partite
information~\cite{Kitaev:2005dm}
\begin{align}\label{}
  I(A,B,C)
  &:=S(A)+S(B)+S(C) -S(AB) \nonumber \\
  &\hspace{1cm}-S(AC)-S(BC)+S(ABC)
\end{align}
does not vanish but rather presents a global indicator for the boundary conditions, in the
sense that it does not depend on the specific layout of the intervals.

\section{Relative entropy}
\label{Relative}
Another essential concept in information theory is \emph{relative entropy}, also known
as the Kullback-Leibler divergence. It is a measure of the distinguishability between two
probability distributions $\rho$ and $\sigma$. Although it is not symmetric, it has a
number of properties that make it fundamental due to its connection to many other
information functions. It is defined by
\[
  D(\rho|\sigma)= \Tr [\rho \log \rho] - \Tr [\rho \log \sigma].
\]

Relative entropy is always positive and vanishes only if $\rho=\sigma$. As is well known,
it can also be rewritten as
\begin{equation}
  \label{dK-dS}
  D(\rho|\sigma) = \Delta \langle K\rangle - \Delta S
\end{equation}
in terms of the variations of the modular Hamiltonian
$K_\sigma=-\log \sigma - \log Z_\sigma$ and the entanglement entropy
\begin{align*}
  \Delta \langle K \rangle &= \langle K_\sigma\rangle_\rho - \langle K_\sigma\rangle_\sigma \\
  \Delta S &= S(\rho)-S(\sigma).
\end{align*}

Here, $\sigma$ is the reference state since we used the modular Hamiltonian associated to
$\sigma$ instead of $\rho$. The convenience of~\eqref{dK-dS} is that presented in this
form it closely resembles the first law of thermodynamics. Given that we have both the
modular Hamiltonian from~\cite{Fries:2019ozf} and the entanglement entropy from
section~\ref{EE}, we are in position to compute the relative entropy directly
using~\eqref{dK-dS}.

Consider the following class of density distributions. Given a global thermal state at
some modulus $\tau$, the reduced density matrix on the subregion $A$ provides a reference
state $\sigma=f(\tau)$. If we now start from a global state at some different modulus
$\tau'$ and reduce to the same region $A$, this will produce another state
$\rho=f(\tau')$. We will consider the relative entropy between two such states.

We start by reviewing the structure of the modular Hamiltonian and then computing its
expectation value. As referred to in the introduction, an explicit expression of the
modular Hamiltonian for chiral fermions on the torus was found in~\cite{Fries:2019ozf}. It
contains a local and a bi-local term:
\[
  K=K^{\text{loc}}+K^{\text{bi-loc}}
\]

As an operator, the local part is spin-independent and takes the standard geometric form,
\begin{equation}
  \label{Kloc}
  K^{\text{loc}} = \int_A \dif x \beta(x) T(x)
\end{equation}
where the stress-tensor of the fermion is
\begin{equation}
  \label{T}
  T(x)=\frac{\ii}{2} \Big[\psi^\dag\partial_x \psi -\psi \partial \psi^\dag \Big](x)
\end{equation}
and $\beta(x)$ is known as the \emph{entanglement temperature}, given by
\[
  \beta(x) = \frac{2\pi \beta}{2\pi +\beta \partial_x \log \Omega(x|\tau)}.
\]

The entanglement temperature is the natural generalisation of the more familiar
Unruh temperature measured by an accelerated observer in the vacuum. Close to the
endpoints of each interval, $\beta(x) \sim x$ and the modular Hamiltonian resembles that of
Rindler space.

The bi-local term was the novel feature found for fermions on the torus. It couples a
given point $x$ to an infinite but discrete set of other points $x_k(x)$ on the entangling
region via
\[
  K^{\text{bi-loc}}_\pm = \int_A \dif x \sum_{k\in \zz}(\pm 1)^k \tilde \beta(x,x_k(x)) \psi^\dag(x)\psi(x_k(x)),
\]
where $+,-$ correspond to the spin sectors $\nu=2,3$, respecively. Here, the points $x_k$
are solutions to the transcendental equation
\begin{equation}\label{xk}
  x-x_k+ \beta g(x,x_k)-k=0
\end{equation}
with
\[
  g(x,y)=\frac{1}{2\pi L} \log \frac{\Omega(x|\tau)}{\Omega(y|\tau)}
\]
and $\Omega(z|\tau)$ as defined in~\eqref{Omega}. The bi-local entanglement temperature is
\[
  \tilde \beta(x,y) = \frac{\ii\pi}{L \left( 1+\beta \partial_x g(x,y) \right) \sinh \pi g(x,y)}.
\]

Now, we consider the expectation value of each term in the modular Hamiltonian, which we
will eventually replace back into the relative entropy~\eqref{dK-dS}. We start with the
local term~\eqref{Kloc}. Notice that the expectation value acts only on the operator $T$
in~\eqref{T}, while the entanglement temperature $\beta(x)$ is fixed by the reference
state $\sigma$ (at modulus $\tau$) and remains the same for the perturbed state $\rho$. As
expected, $\langle T\rangle$ is divergent due to operators evaluated at identical
space-time positions, but relative entropy is well defined since these UV divergences
cancel in~\eqref{dK-dS}. Using the shorthand notation
\begin{equation}
  \label{DG}
  \Delta G_\nu(z) = G_\nu(z|\tau')-G_\nu(z|\tau)
\end{equation}
the variation of the local term reads
\begin{equation}
  \label{dKloc}
  \Delta \langle K^{\text{loc}}\rangle =-\ii \partial_z \Delta G_\nu(z)\Big|_{z=0} \int_A \dif x \beta(x).
\end{equation}

Although we have not found a closed expression for the integral, it is easy to compute it
numerically. Notice that, while $K^{\text{loc}}$ as an operator is independent of the spin
sector, its expectation value is not. Indeed, the prefactor depends on the derivative of
the propagator $G_\nu$. As we comment in the discussion, this implies that relative
entropy is not spin-independent.

On the other hand, the variation of the bi-local contribution is
\begin{align}
  \label{dKbiloc}
  \Delta \langle K^{\text{bi-loc}}\rangle
  &= \sum_{k\in\zz} \int_A \dif x (\pm 1)^k \tilde\beta(x,x_k(x)) \nonumber \\
  &\hspace{2.5cm} \times \Delta G_\nu(x-x_k(x)).
\end{align}
This integral is technically difficult to deal with: one must first determine the
solutions $x_k(x)$ to~\eqref{xk}---a transcendental equation involving elliptic
functions---and then perform the integration.

Finally, the variation of the entanglement entropy also has two contributions,
\[
  \Delta S=S(\tau')-S(\tau).
\]
Putting everything together, the relative entropy is given by:
\begin{equation}
  \label{Dfinal}
  D(\rho|\sigma) =
  \Delta \langle K^{\text{loc}}\rangle + \Delta \langle K^{\text{bi-loc}}\rangle - \Delta S^{(0)} - \Delta S^{(\nu)},
\end{equation}
where each term is given in~\eqref{dKloc}, \eqref{dKbiloc}, \eqref{S0CH}, \eqref{S3},
and~\eqref{S2}. As emphasised above, in practice the evaluation of~\eqref{Dfinal} is very
challenging and we leave it as future work.

\subsection{An example: the thermal cylinder}
In order to illustrate how to compute in principle this relative entropy, let us simplify
things and perform the explicit calculation for a single interval of length $ L$, in the
limit when the size of the spatial circle is much larger than both the entangling region
and the temporal cycle. We view this as an exercise combining known tools rather than a
new result. Then, our expressions for the modular Hamiltonian and entanglement entropy on
the torus reduce to the well known universal result on the thermal
cylinder~\cite{Cardy:2016fqc}
\begin{align*}
  K(\beta)
  &= \beta \int_0^ L \dif x \frac{\sinh \frac{\pi(L-x)}\beta \sinh\frac{\pi x}\beta}{\sinh\frac{\pi L}\beta} T(x),   \\
  S(\beta)
  &= \frac{1}{6}\log \bigg[\frac{\beta}{\pi \epsilon} \sinh  \frac{\pi  L}{\beta}  \bigg],
\end{align*}
where $T(x)$ is again the fermionic stress tensor. Notice that in this case the modular
flow is purely local. As described above, by varying the inverse temperature $\beta$ of
the parent state, we get a one-parameter family of states defined on $A$. We can compute
the relative entropies between such states.

\begin{figure}[h]
\begin{center}
\includegraphics[width=8cm]{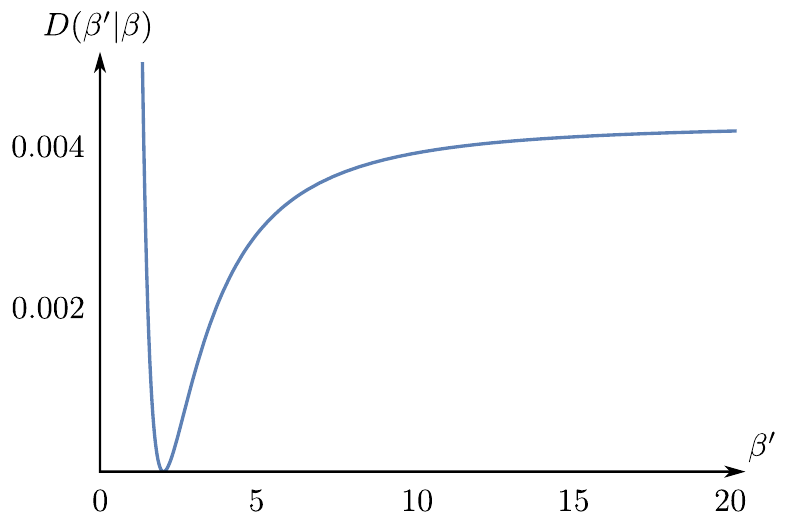} 
\end{center}
\caption{Relative entropy between the states obtained by reducing a thermal state on the
  line at temperature $\beta'$ and $\beta$ respectively, to a single interval of length
  $ L=1$. The reference temperature associated to $\sigma$ is $\beta=2$.}
\label{figrelative}
\end{figure}

For this simplified case it is easy to obtain~\eqref{dKloc} from the fermionic propagator,
and find an explicit expression for the relative entropy:
\begin{align*}
  D(\beta|\beta')
  &=\frac{\beta}{24} \bigg[\frac 1{\beta'^2} - \frac 1{\beta^2}\bigg] \bigg[-\beta +\pi  L \coth \frac{\pi L}\beta\bigg] \\
  &\hspace{3.5cm} +\frac 16 \log \frac{\beta \sinh \frac{\pi L}\beta}{\beta' \sinh \frac{\pi L}{\beta'}}.
\end{align*}
Finally, it is illustrative to plot this function for a fixed reference $\beta=2$ and
interval length $ L=1$, while we vary $\beta'$, see Fig.~\ref{figrelative}.

As can be seen from the figure, the distribution obtained from a parent state at low
$\beta'$ (high temperature) is very distinguishable from the finite temperature state. On
the other hand, the relative entropy with respect to the vacuum $\beta'\to\infty$
asymptotes to a constant. Relative entropy only vanishes at $\beta'=\beta$, as it should,
and the fact that the curve has vanishing slope at that point is the \emph{first law of
  entanglement}.  Again, we take this as a pedagogical first step in the explicit
computation of relative entropy, and leave for future work the more involved calculations
on the torus.

\section{The resolvent with a chemical potential}
\label{ChemicalPotential}
In order to repeat the calculations for the two remaining cases $\nu = 1,4$ (R-R
and NS-R), let us derive the resolvent in these sectors. We will do this explicitly for
$\nu = 1$, the other case is entirely analogous. Consider first the Green's function
\begin{equation}
  \label{eq:g-chem-series}
  G(z|\tau,\mu)
  = \sum_{k \in \zz} \frac{\ee^{-2\pi\ii k z}}{1-\ee^{2\pi (\ii k\tau-\mu)}},
\end{equation}
where we introduced a \emph{chemical potential} $\mu$ to regularize the divergent term at
$k=0$. The above series does not converge pointwise, however, it does converge in the
sense of distributions: Starting from the geometric series
\[
  \frac 1{2\ii} \cot \pi z = \frac 12
  \frac{\ee^{\ii \pi z}+\ee^{-\ii \pi z}}{\ee^{\ii \pi z}-\ee^{-\ii \pi z}} =
  \frac 12 + \sum_{k \geq 1}\ee^{-2 \pi\ii k z}
\]
for $\Im[z] < 0$, we find
\begin{align*}
  &G(z|\tau,\mu) = \frac 1{2\ii} \cot \pi z - \frac 12 + \frac 1{1-\ee^{-2\pi \mu}} \\
  &\quad+ \sum_{\substack{k \geq 2 \\ \text{even}}} \Big[\frac{\ee^{\ii\pi k z}}{1-\ee^{-2\pi\mu}q^{-k}}
    + \frac{\ee^{-\ii\pi k z}}{1-\ee^{-2\pi\mu}q^k} - \ee^{-\ii\pi k z}\Big],
\end{align*}
which can be rewritten as
\begin{align*}
  G(z|\tau,\mu)
  &= \frac 1{2\ii} \cot \pi z + \frac 12 \frac{1+\ee^{-2\pi \mu}}{1-\ee^{-2\pi \mu}}\\
  &\quad-\sum_{\substack{k \geq 2 \\ \text{even}}}
    \bigg[\frac{\ee^{\ii\pi k z}q^k}{\ee^{-2\pi\mu}-q^k}-\frac{\ee^{-\ii\pi k z}q^k}{\ee^{2\pi \mu}-q^k}\bigg].
\end{align*}
The right hand side is now absolutely convergent on the strip $-\Im[\tau] <
\Im[z] < \Im[\tau]$ and yields
\begin{equation}
  \label{eq:g-chem}
  G_\nu(z|\tau,\mu) = \frac{\eta^3(\tau)}{\ii \vartheta_1(z|\tau)}
  \frac{\vartheta_\nu(z-\ii\mu|\tau)}{\vartheta_\nu(-\ii\mu|\tau)}
\end{equation}
for general spin structure $\nu=1,2,3,4$.

As can be readily seen from the quasiperiodicities of the theta functions,
eq.~\eqref{eq:g-chem} satisfies the generalized KMS condition
\begin{equation}
  \label{eq:gen-kms}
  G(z+\tau|\tau,\mu) = \pm \ee^{-2\pi \mu} G(z|\tau,\mu).
\end{equation}
We can thus use the methodology of~\cite{Fries:2019ozf} to derive the resolvent
as~\eqref{R} with
\begin{equation}
  \label{eq:f-chem}
  F(x,y;\xi) =\frac{\xi-1/2}{\xi+1/2}
  \bigg[\frac{\Omega(x)}{\Omega(y)}\bigg]^{\ii h} \!\! G(x-y|\tau,\mu+Lh).
\end{equation}
Note that, unlike the Green's function~\eqref{eq:g-chem}, the resolvent is well defined in
the limit $\mu \to 0$.

Having found the resolvent, we can now in principle reapeat our calculations for the cases
$\nu=1,4$. However, there is one caveat: Because of the bose statistics
in~\eqref{eq:g-chem-series}, $G$ has an unbounded spectrum and thus we have to find new
valid formulae for the modular Hamiltonian, entanglement, and relative entropy in terms of
the resolvent. While this is easy for the modular Hamiltonian, we have yet to find good
expressions for the entropies.

\section{Discussion}
\label{Discussion}
In this short note we have described some applications of the modular Hamiltonian found
in~\cite{Fries:2019ozf} for the chiral fermion on the circle at finite temperature. First,
we found an exact expression for the entanglement entropy for multiple intervals, valid
for any torus modulus $\tau$, i.e. generic spatial and thermal circles. This is a
generalisation of the results in~\cite{Herzog:2013py,Azeyanagi:2007bj}, where the authors
provided the entanglement entropy in the high and low temperature expansions. Whereas the
spin-independent contribution to the entanglement entropy was known from these works, the
novel result of this paper was to derive the analytic continuation of the spin-dependent
term for an arbitrary torus.

Remarkably, this piece depends on the entangling intervals only via the total length of
the region $L=\sum_j |b_j-a_j|$, and not the details of its configuration. This implies
that the three-partite information of three intervals is a, generally non-vanishing,
global indicator of the spin structure.

The third result of this paper is an explicit formula for the relative entropy
$D(\rho|\sigma)$ of the free fermion on the torus. As the reference state, we chose the
reduced density matrix $\sigma$ on multiple intervals coming from a global state on the
torus of modulus $\tau=i\beta$. The target state is obtained in the same way but starting
from a different temperature $\beta'$. This provides a one-parameter family of states for
computing relative entropy. Although in practice the integrals involved in the computation
are very challenging, mainly due to the non-local terms involved, the final
expression~\eqref{Dfinal} is explicit and can be investigated numerically.

Moreover, we also observed that while both the modular Hamiltonian as an operator and the
entanglement entropy separate into a universal and spin-dependent parts, this does not
hold for the relative entropy. This is simply because the computation of relative entropy
involves expectation values which yields Green's functions in the modular
Hamiltonians. Since relative entropy is a measure of the distinguishability of two
distributions, this implies that fermions with some boundary conditions are more
distinguishable than others.

Finally, we showed that the propagator of the fermion on the torus with a chemical
potential allows to compute the resolvent for the spin sectors $\nu=1,4$ which were not
considered in~\cite{Fries:2019ozf}. Moreover, the propagator-like term that appears in the
resolvent corresponds precisely to the propagator with a chemical potential. This is
due to the identical boundary conditions imposed in both problems.

\section{Acknowledgments}
PF is financially supported by the DFG project DFG HI 744/9-1. The work of IR is funded by
the Gravity, Quantum Fields and Information group at AEI, which is generously supported by
the Alexander von Humboldt Foundation and the Federal Ministry for Education and Research
through the Sofja Kovalevskaja Award. IR also acknowledges the hospitality of Perimeter
Institute, where part of this work was done.

\begin{appendix}
\section{Conventions}
Our definitions for Dedekind eta function is
\[
  \eta(\tau) := e^{\pi i \tau/12} \prod_{k\geq 1} (1-q^{2k})
\]
in terms of the nome
\[
  q := \ee^{\ii\pi\tau}.
\]
Furthermore, we define the Jacobi theta function by
\[
  \vartheta_3(z|\tau) := \sum_{k\in\zz} q^{n^2} e^{2\pi \ii z},
\]
while the auxiliary thetas are given by
\begin{align*}
  \vartheta_4(z|\tau) &= \vartheta_3(z+1/2|\tau), \\
  \vartheta_2(z|\tau) &= q^{1/4}\ee^{\ii\pi z} \vartheta_3(z+\tau/2|\tau), \\
  \vartheta_1(z|\tau) &= -\ii q^{1/4} \ee^{\ii\pi z} \vartheta_4(z+1/2+\tau/2|\tau).
\end{align*}

\section{Laurent series of the propagator}
The propagators appearing in~\eqref{Gzmu} are given by the following Laurent series,
\begin{align}
  G_3(z|\tau,Lh)
  &=\frac{1}{2\ii\sin \pi z} \nonumber \\
  &\hspace{1cm}+ \sum_{\substack{k\geq 1\\ \text{odd}}} \bigg[\frac{\ee^{\ii\pi k}q^k}{\Lambda^{-1}+q^k}-\frac{\ee^{-\ii\pi k}q^k}{\Lambda+q^k} \bigg] \label{Laurent3} \\
  G_2(z|\tau,Lh)
  &=\frac{1}{2\ii}\cot \pi z + \frac 12 \frac{\Lambda-1}{\Lambda+1} \nonumber \\
  &\hspace{1cm}+ \sum_{\substack{k\geq 2\\ \text{even}}} \bigg[\frac{\ee^{\ii\pi k}q^k}{\Lambda^{-1}+q^k}-\frac{\ee^{-\ii\pi k}q^k}{\Lambda+q^k} \bigg], \label{Laurent2}
\end{align}
where again $\Lambda=e^{2 \pi L h}$.

\end{appendix}


\end{document}